\documentclass{ws-procs975x65}%
\usepackage{amsmath}
\usepackage{amsfonts}
\usepackage{amssymb}
\usepackage{graphicx}%
\def\beq{\begin{equation}}
\def\eeq{\end{equation}}
\begin{document}

\title{Black Hole Thermodynamics and Gravity's Rainbow}
\author{Remo Garattini}

\address{Universit\`{a} degli Studi di Bergamo, \\
Dipartimento di Ingegneria e Scienze Applicate,\\
Viale Marconi,5 24044 Dalmine (Bergamo) ITALY\\
I.N.F.N. - sezione di Milano, Milan, Italy\\
$^*$E-mail:remo.garattini@unibg.it}

\begin{abstract}
We consider the effects of rotations on the calculation of some
thermodynamical quantities like the free energy, internal energy and entropy.
In ordinary gravity, when we evaluate the density of states of a scalar field
close to a black hole horizon, we obtain a divergent result which can be kept
under control with the help of some standard regularization and
renormalization processes. We show that when we use the Gravity's Rainbow
approach such regularization/renormalization processes can be avoided. A
comparison between the calculation done in an inertial frame and in a comoving
frame is presented.
\end{abstract}

\bodymatter

\section{Introduction}

\label{p1}Gravity's Rainbow (GRw) is a modification of the space-time close to
the Planck scale. It has been introduced for the first time by Magueijo and
Smolin\cite{MagSmo}. Basically one defines two unknown functions $g_{1}\left(
E/E_{P}\right)  $ and $g_{2}\left(  E/E_{P}\right)  $ having the following
property%
\begin{equation}
\lim_{E/E_{P}\rightarrow0}g_{1}\left(  E/E_{P}\right)  =1\qquad\text{and}%
\qquad\lim_{E/E_{P}\rightarrow0}g_{2}\left(  E/E_{P}\right)  =1.
\end{equation}
This property guarantees the recovery of ordinary General Relativity when
sub-planckian physics is involved. In this formalism, the Einstein's field
equations are replaced by a one parameter family of equations%
\begin{equation}
G_{\mu\nu}\left(  E\right)  =8\pi G\left(  E\right)  T_{\mu\nu}\left(
E\right)  +g_{\mu\nu}\Lambda\left(  E\right)  ,\label{EEM}%
\end{equation}
where $G\left(  E\right)  $ is an energy dependent Newton's constant and
$\Lambda\left(  E\right)  $ is an energy dependent cosmological constant,
respectively. They are defined so that $G\left(  0\right)  $ is the physical
Newton's constant and $\Lambda\left(  0\right)  $ is the usual cosmological
constant. In this context, the Schwarzschild solution of $\left(
\ref{EEM}\right)  $ becomes%
\begin{equation}
ds^{2}\left(  E\right)  =-\left(  1-\frac{2MG\left(  0\right)  }{r}\right)
\frac{dt^{2}}{g_{1}^{2}\left(  E/E_{P}\right)  }+\frac{dr^{2}}{\left(
1-\frac{2MG\left(  0\right)  }{r}\right)  g_{2}^{2}\left(  E/E_{P}\right)
}+\frac{r^{2}}{g_{2}^{2}\left(  E/E_{P}\right)  }d\Omega^{2}.\label{ds}%
\end{equation}
An immediate generalization of the metric $\left(  \ref{ds}\right)  $ is
represented by the following line element%
\begin{equation}
ds^{2}\left(  E\right)  =-\left(  1-\frac{b\left(  r\right)  }{r}\right)
\frac{\exp\left(  -2\Phi\left(  r\right)  \right)  }{g_{1}^{2}\left(
E/E_{P}\right)  }dt^{2}+\frac{dr^{2}}{\left(  1-\frac{b\left(  r\right)  }%
{r}\right)  g_{2}^{2}\left(  E/E_{P}\right)  }+\frac{r^{2}}{g_{2}^{2}\left(
E/E_{P}\right)  }d\Omega^{2}.\label{ds1}%
\end{equation}
The function $b\left(  r\right)  $ will be referred to as the
\textquotedblleft shape function\textquotedblright\ and it may be thought of
as specifying the shape of the spatial slices. The location of the horizon is
determined by the equation $b\left(  r_{H}\right)  =r_{H}$. On the other hand,
$\Phi\left(  r\right)  $ will be referred to as the \textquotedblleft redshift
function\textquotedblright\ and describes how far the total gravitational
redshift deviates from that implied by the shape function. The line element
$\left(  \ref{ds1}\right)  $ describes any spherically symmetric space-time
with a horizon: by definition, this is a black hole distorted by GRw. Note
that a metric of the form%
\begin{equation}
ds^{2}\left(  E\right)  =-\frac{\exp\left(  -2\Phi\left(  r\right)  \right)
}{g_{1}^{2}\left(  E/E_{P}\right)  }dt^{2}+\frac{dr^{2}}{\left(
1-\frac{b\left(  r\right)  }{r}\right)  g_{2}^{2}\left(  E/E_{P}\right)
}+\frac{r^{2}}{g_{2}^{2}\left(  E/E_{P}\right)  }d\Omega^{2},\label{ds2}%
\end{equation}
describes a traversable wormhole modified by GRw if $\exp\left(  -2\Phi\left(
r\right)  \right)  $ never vanishes\cite{Visser}. The line element $\left(
\ref{ds2}\right)  $ has been used in a series of papers where to avoid any
regularization/renormalization scheme which appear in conventional Quantum
Field Theory calculations like one loop corrections to classical
quantities\cite{Remo}. On the other hand, the line element $\left(
\ref{ds2}\right)  $ has been considered for the computation of black hole
entropy\cite{RemoPLB}. In this last case, the idea is to avoid to introduce a
cut-off of Planckian size known as \textquotedblleft\textit{brick
wall}\textquotedblright\cite{tHooft}. The \textquotedblleft\textit{brick
wall}\textquotedblright\ appears when one uses a statistical mechanical
approach to explain the famous Bekenstein-Hawking
formula\cite{Bekenstein,Hawking}%
\begin{equation}
S_{BH}=\frac{1}{4}A/l_{P}^{2},
\end{equation}
relating the entropy of a black hole and its area. Indeed, when one tries to
adopt such an approach, one realizes that the density of energy levels of
single-particle excitations is divergent near the horizon. Of course, several
attempts have been done to avoid the introduction of the \textit{brick wall}.
For instance, without modifying gravity at any scale, it has been suggested
that the \textit{brick wall} could be absorbed in a renormalization of
Newton's constant\cite{SusUgl,BarEmp,EWin}, while other authors approached the
problem of the divergent brick wall using Pauli-Villars
regularization\cite{DLM,FS,KKSY}. Other than GRw other proposals have been
made in the context of modified gravity. For instance, non-commutative
geometry introduces a natural thickness of the horizon replacing the 't
Hooft's brick wall\cite{BaiYan} and \textit{Generalized Uncertainty Principle}
(GUP) modifies the Liouville measure\cite{XL,RenQinChun,GAC,VAHA}.
Nevertheless we can wonder what happens when one introduces rotations. For
instance, one could consider the free energy obtained for a real massless
scalar field, rotating with an angular velocity $\Omega_{0}$ around the $z$
axis in Minkowski space%
\begin{equation}
F=\frac{1}{\beta}\sum_{m}\int_{0}^{\infty}dg\left(  E,m\right)  \ln\left(
1-e^{-\beta\left(  E-m\Omega_{0}\right)  }\right)  .\label{FRot}%
\end{equation}
For this case, it is better to work in cylindrical coordinates%
\begin{equation}
ds^{2}=-dt^{2}+dr^{2}+r^{2}d\phi^{2}+dz^{2}\label{Flat}%
\end{equation}
and with the help of a WKB approximation it is possible to show that a
divergence appear close to the speed-of-light (SOL) surface, defined as the
surface where $r=\Omega_{0}^{-1}$ \cite{LK}. This divergence can be taken
under control with the help of GRw by modifying the line element $\left(
\ref{Flat}\right)  $ in the following way\cite{RemoRot}%
\begin{equation}
ds^{2}=-\frac{dt^{2}}{g_{1}^{2}\left(  E/E_{P}\right)  }+\frac{dr^{2}}%
{g_{2}^{2}\left(  E/E_{P}\right)  }+\frac{r^{2}d\phi^{2}}{g_{2}^{2}\left(
E/E_{P}\right)  }+\frac{dz^{2}}{g_{2}^{2}\left(  E/E_{P}\right)  },\label{ds0}%
\end{equation}
where for simplicity we have replaced $ds^{2}\left(  E\right)  $ with $ds^{2}%
$. The same above thermodynamical system can be analyzed on a comoving frame
rotating with the same angular velocity $\Omega_{0}$. By plugging
$\phi^{\prime}=\phi-\Omega_{0}t$, from the line element $\left(
\ref{ds}\right)  $ we obtain%
\begin{equation}
ds^{2}=-\frac{\left(  1-\Omega_{0}^{2}r^{2}\right)  }{g_{1}^{2}\left(
E/E_{P}\right)  }dt^{2}+\frac{2\Omega_{0}r^{2}d\phi^{\prime}dt}{g_{1}\left(
E/E_{P}\right)  g_{2}\left(  E/E_{P}\right)  }+\frac{dr^{2}}{g_{2}^{2}\left(
E/E_{P}\right)  }+\frac{r^{2}d\phi^{\prime2}}{g_{2}^{2}\left(  E/E_{P}\right)
}+\frac{dz^{2}}{g_{2}^{2}\left(  E/E_{P}\right)  }.\label{dsC}%
\end{equation}
It is immediate to see that in this system of coordinates appears a fictitious
horizon located at $r=\Omega_{0}^{-1}$, namely the SOL surface of the rotating
heat bath introduced in $\left(  \ref{FRot}\right)  $. Note that a mixing
between $g_{1}\left(  E/E_{P}\right)  $ and $g_{2}\left(  E/E_{P}\right)  $
appears. A similar mixing appears also in a Vaidya spacetime for%
\begin{equation}
ds^{2}=-\left(  1-\frac{2M\left(  v\right)  G}{r}\right)  \frac{dv^{2}}%
{g_{1}^{2}\left(  E/E_{P}\right)  }+2\frac{dvdr}{g_{1}\left(  E/E_{P}\right)
g_{2}\left(  E/E_{P}\right)  }+\frac{r^{2}d\Omega^{2}}{g_{2}^{2}\left(
E/E_{P}\right)  },
\end{equation}
where $v$ is the advanced (ingoing) null coordinate and finally also for the
Kerr metric which, in the context of GRw, becomes\cite{ZL}%
\begin{equation}
ds^{2}=\frac{g_{tt}dt^{2}}{g_{1}^{2}\left(  E/E_{P}\right)  }+\frac{2g_{t\phi
}dtd\phi}{g_{1}\left(  E/E_{P}\right)  g_{2}\left(  E/E_{P}\right)  }%
+\frac{g_{\phi\phi}d\phi^{2}}{g_{2}^{2}\left(  E/E_{P}\right)  }+\frac
{g_{rr}dr^{2}}{g_{2}^{2}\left(  E/E_{P}\right)  }+\frac{g_{\theta\theta
}d\theta^{2}}{g_{2}^{2}\left(  E/E_{P}\right)  },\label{e32}%
\end{equation}
where%
\begin{align}
g_{tt} &  =-\frac{\Delta-a^{2}\sin^{2}\theta}{\Sigma},\qquad g_{t\phi}%
=-\frac{a\sin^{2}\theta\left(  r^{2}+a^{2}-\Delta\right)  }{\Sigma
},\nonumber\\
g_{\phi\phi} &  =\frac{\left(  r^{2}+a^{2}\right)  ^{2}-\Delta a^{2}\sin
^{2}\theta}{\Sigma}\sin^{2}\theta,\qquad g_{rr}=\frac{\Sigma}{\Delta},\qquad
g_{\theta\theta}=\Sigma,
\end{align}
and%
\begin{equation}
\Delta=r^{2}-2MGr+a^{2},\qquad\Sigma=r^{2}+a^{2}\cos^{2}\theta.
\end{equation}
Here $M$ and $a$ are mass and angular momentum per unit mass of the black
hole, respectively. $\Delta$ vanishes when $r=r_{\pm}=MG\pm\sqrt{\left(
MG\right)  ^{2}-a^{2}}$, while $g_{tt}$ vanishes when $r=r_{S\pm}=$
$MG\pm\sqrt{\left(  MG\right)  ^{2}-a^{2}\cos^{2}\theta}$: they are not
modified by GRw and the outer horizon or simply horizon is located at
$r_{+}=r_{H}$. Units in which $\hbar=c=k=1$ are used throughout the paper.

\section{GRw Entropy for the Kerr Black Hole}

To discuss the entropy for a Kerr black hole we have two options: we can use a
rest observer at infinity (ROI) or we can use a Zero Angular Momentum Observer
(ZAMO)\cite{CYLY, LK}. The ROI\ frame is described by the line element
$\left(  \ref{e32}\right)  $ and the appropriate form of the free energy is
the following%
\begin{equation}
F=\frac{1}{\beta}\int_{0}^{\infty}dn\left(  E\right)  \ln\left(
1-e^{-\beta\left(  E-m\Omega_{0}\right)  }\right)  .\label{FK}%
\end{equation}
It is immediate to see that when we use a ROI, the problem of superradiance
appears when the free energy $\left(  \ref{FK}\right)  $ is computed in the
range $0<E<m\Omega_{0}$. On the other hand when a ZAMO is considered, the free
energy $\left(  \ref{FK}\right)  $ becomes similar to the one used for a
Schwarzschild black hole. Basically this happens because near the horizon the
metric becomes
\begin{equation}
ds^{2}=-\frac{N^{2}dt^{2}}{g_{1}^{2}\left(  E/E_{P}\right)  }+g_{\phi\phi
}\frac{d\phi^{2}}{g_{2}^{2}\left(  E/E_{P}\right)  }+g_{rr}\frac{dr^{2}}%
{g_{2}^{2}\left(  E/E_{P}\right)  }+g_{\theta\theta}\frac{d\theta^{2}}%
{g_{2}^{2}\left(  E/E_{P}\right)  }\label{e32c}%
\end{equation}
and the mixing between $t$ and $\phi$ disappears. Moreover when we use a ZAMO
frame, the superradiance does not come into play because there is no
ergoregion. Indeed since we have defined%
\begin{equation}
N^{2}=g_{tt}-\frac{g_{t\phi}^{2}}{g_{\phi\phi}}=-\frac{1}{g^{tt}}%
=-\frac{\Delta\sin^{2}\theta}{g_{\phi\phi}},
\end{equation}
$N^{2}$ vanishes when $r\rightarrow r_{H}$. The number of modes with frequency
less than $E$ is given approximately by
\begin{equation}
n(E)=\frac{1}{\pi}\int_{0}^{l_{max}}(2l+1)\int_{r_{H}}^{R}\sqrt{k^{2}%
(r,l,E)}drdl,
\end{equation}
Here it is understood that the integration with respect to $r$ and $l$ is
taken over those values which satisfy $r_{H}\leq r\leq R$ and $k^{2}%
(r,l,E)\geq0$. Thus one finds%
\begin{equation}
\frac{dn(E)}{dE}=\frac{1}{8\pi^{2}}\int d\theta d\bar{\phi}\int_{r_{H}}%
^{R}dr\left(  -g^{tt}\right)  ^{\frac{3}{2}}\sqrt{g_{rr}g_{\theta\theta
}g_{\phi\phi}}\frac{1}{3}\frac{d}{dE}\left(  h^{3}\left(  E/E_{P}\right)
E^{3}\right)  .
\end{equation}
In proximity of the horizon, the free energy can be approximated by%
\begin{equation}
F_{r_{H}}=\frac{1}{8\pi^{2}\beta}\int d\theta d\bar{\phi}\int_{0}^{\infty}%
\ln\left(  1-e^{-\beta E}\right)  \frac{d}{dE}\left(  \frac{1}{3}h^{3}\left(
E/E_{P}\right)  E^{3}\right)  H\left(  r_{H},r_{1}\right)  dE
\end{equation}
where we have defined%
\begin{equation}
H\left(  r_{H},r_{1}\right)  =\int_{r_{H}}^{r_{1}}dr\left(  -g^{tt}\right)
^{\frac{3}{2}}\sqrt{g_{rr}g_{\theta\theta}g_{\phi\phi}}.
\end{equation}
$F_{r_{H}}$ can be further reduced to%
\begin{equation}
F_{r_{H}}\simeq\frac{C\left(  r_{H},\theta\right)  }{8\pi^{2}\beta}\int
_{0}^{\infty}\frac{\ln\left(  1-e^{-\beta E}\right)  }{\sigma\left(
E/E_{P}\right)  }\frac{d}{dE}\left(  \frac{1}{3}h^{3}\left(  E/E_{P}\right)
E^{3}\right)  dE,
\end{equation}
where%
\begin{equation}
C\left(  r_{H},\theta\right)  =\int d\theta d\bar{\phi}\left[  \frac{\left(
r_{H}^{2}+a^{2}\right)  ^{4}\sin\theta}{r_{H}\left(  r_{H}-r_{-}\right)
^{2}\Sigma_{H}}\right]
\end{equation}
and where we have assumed that, in proximity of the throat the brick wall can
be written as $r_{0}\left(  E/E_{P}\right)  =r_{H}\sigma\left(  E/E_{P}%
\right)  $ with%
\begin{equation}
\sigma\left(  E/E_{P}\right)  \rightarrow0,\qquad E/E_{P}\rightarrow0.
\end{equation}
With an integration by parts one finds%
\begin{align}
F_{r_{H}}  & =-\frac{C\left(  r_{H},\theta\right)  }{24\pi^{2}\beta}\int
_{0}^{\infty}\frac{E^{3}h^{3}\left(  E/E_{P}\right)  }{\sigma\left(
E/E_{P}\right)  }\nonumber\\
& \times\left[  \frac{\beta}{\left(  \exp\left(  \beta E\right)  -1\right)
}-\frac{\ln\left(  1-e^{-\beta E}\right)  }{E_{P}\sigma\left(  E/E_{P}\right)
}\sigma^{\prime}\left(  E/E_{P}\right)  \right]  dE.
\end{align}
It is possible to show that%
\begin{align}
F_{r_{H}} &  =-\frac{C\left(  r_{H},\theta\right)  }{24\pi^{2}\beta}\int
_{0}^{\infty}\left[  \frac{\beta Ee^{-3E/E_{P}}}{\left(  \exp\left(  \beta
E\right)  -1\right)  }-2e^{-3E/E_{P}}\ln\left(  1-e^{-\beta E}\right)
\right]  dE\nonumber\\
&  =-\frac{C\left(  r_{H},\theta\right)  }{24\pi^{2}\beta}\left[  \zeta\left(
2,1+\frac{3}{\beta E_{P}}\right)  +\frac{\beta E_{P}}{3}\left(  \gamma
+\Psi\left(  1+\frac{3}{\beta E_{P}}\right)  \right)  \right]  ,
\end{align}
where $\zeta\left(  s,\nu\right)  $ is the Hurwitz zeta function,
$\Gamma\left(  x\right)  $ is the gamma function and $\Psi\left(  x\right)  $
is the digamma function. In the limit where $\beta E_{P}\gg1$, at the leading
order, one finds that the entropy can be approximated by%
\begin{equation}
S=\beta^{2}\frac{\partial F_{r_{w}}}{\partial\beta}=\frac{E_{P}^{2}}{36\beta
}\int d\theta d\bar{\phi}\left[  \frac{\left(  r_{H}^{2}+a^{2}\right)
^{4}\sin\theta}{r_{H}\left(  r_{H}-r_{-}\right)  ^{2}\Sigma_{H}}\right]
\label{S}%
\end{equation}
and even when rotation is included, the \textquotedblleft\textit{brick
wall}\textquotedblright\ does not appear. Of course the entropy $\left(
\ref{S}\right)  $ can always be cast in the familiar form%
\begin{equation}
S=\frac{A_{H}}{4G},
\end{equation}
where $A_{H}$ is the horizon area. To summarize, we have shown that the
ability of Gravity's Rainbow to keep under control the UV divergences applies
also to rotations. However the connection between a ROI and a ZAMO has to be
investigated with care\cite{Kerr}. Indeed in the ROI frame, the superradiance
phenomenon appears, while in the ZAMO frame does not. Once the connection is
established nothing forbids to extend this result to other rotating
configuration like, for example, Kerr-Newman or Kerr-Newman-De Sitter (Anti-De Sitter).

\section*{Acknowledgments}

The author would like to thank MDPI for a partial financial support.

\end{document}